\documentclass[conference]{IEEEtran}
\makeatletter
\def\endthebibliography{%
  \def\@noitemerr{\@latex@warning{Empty `thebibliography' environment}}%
  \endlist
}
\makeatother

\usepackage{cite}
\usepackage{amsmath,amssymb,amsfonts}
\usepackage{algorithmic}
\usepackage{graphicx}
\usepackage{textcomp}
\usepackage{xcolor}
\def\BibTeX{{\rm B\kern-.05em{\sc i\kern-.025em b}\kern-.08em
    T\kern-.1667em\lower.7ex\hbox{E}\kern-.125emX}}

\usepackage[utf8]{inputenc} 
\usepackage[colorlinks]{hyperref}
\usepackage{float}
\usepackage{wrapfig}
\usepackage{bm}
\usepackage{mathtools,cuted}
\usepackage{booktabs} 
\usepackage{array} 
\usepackage{paralist} 
\usepackage{verbatim} 
\usepackage{subcaption} 
\newcommand{\prob}{\mathrm{pr}}
\newcommand{\Prob}{\mathrm{Pr}}

\newcommand{\ent}{\mathrm{H}}
\newcommand{\smi}{\mathrm{I}}
\newcommand{\pe}{\mathrm{P_e}}

\DeclarePairedDelimiterX{\infdivx}[2]{(}{)}{%
  #1||#2%
}
\newcommand{\kld}{\ensuremath{\mathrm{KL}\infdivx}}

\newcommand{\ca}{\ensuremath{\mathrm{C_\alpha}\infdivx}}

\begin{document}

\title{Bounds on mutual information of mixture data for classification tasks}
\author{\IEEEauthorblockN{Yijun Ding}
\IEEEauthorblockA{\textit{James C. Wyant College of Optical Sciences} \\
\textit{University of Arizona}\\
Tucson, AZ, USA \\
dingy@email.arizona.edu}
\and
\IEEEauthorblockN{Amit Ashok}
\IEEEauthorblockA{\textit{James C. Wyant College of Optical Sciences} \\
\textit{and Department of Electrical Engineering}\\
\textit{University of Arizona}\\
Tucson, AZ, USA \\
ashoka@email.arizona.edu}
}

\maketitle
\begin{abstract}
The data for many classification problems, such as pattern and speech recognition, follow mixture distributions. To quantify the optimum performance for classification tasks, the Shannon mutual information is a natural information-theoretic metric, as it is directly related to the probability of error. The mutual information between mixture data and the class label does not have an analytical expression, nor any efficient computational algorithms. We introduce a variational upper bound, a lower bound, and three estimators, all employing pair-wise divergences between mixture components. We compare the new bounds and estimators with Monte Carlo stochastic sampling and bounds derived from entropy bounds. To conclude, we evaluate the performance of the bounds and estimators through numerical simulations. 
\end{abstract}


\section{Introduction}
\subsection{Motivation}
\indent We study the performance of classification tasks, where the goal is to infer the class label $C$ from sample data $\bm x$. The Shannon mutual information $\smi(\bm x; C)$ characterizes the reduction in the uncertainty of the class label $C$ with the knowledge of data $\bm x$ and provides a way to quantify the relevance of the data $\bm x$ with respect to the class label $C$. As the mutual information is related to probability of classification error ($\pe$) through Fano's inequality and other bounds \cite{fano1961transmission, kovalevskij1967problem,hu2016optimization}, it has been widely used for feature selection \cite{vergara2014review,battiti1994using}, learning \cite{tishby2000information}, and quantifying task-specific information \cite{neifeld2007task} for classification.

\indent Statistical mixture distributions such as Poisson, Wishart or Gaussian mixtures are frequently used in the fields of speech recognition \cite{hershey2007approximating}, image retrieval \cite{goldberger2003efficient}, system evaluation \cite{ding2020x}, compressive sensing \cite{duarte2012task}, distributed state estimation \cite{noack2014nonlinear}, hierarchical clustering \cite{goldberger2005hierarchical} etc. As a practical example, consider a scenario in which the data $\bm x$ is measured with a noisy system, eg. Poisson noise in an photon-starved imaging system or Gaussian noise in a thermometer reading. If the actual scene (or temperature) has a class label, e.g. target present or not (the temperature is below freezing or not), then the mutual information $\smi(\bm x; C)$ describes with what confidence one can assign a class label $C$ to the noisy measurement data $\bm x$.

\indent The goal of this paper is to develop efficient methods to quantify the optimum performance of classification tasks, when the distribution of the data $\bm x$ for each given class label, $\prob(\bm x|C)$, follows a known mixture distribution. As the mutual information $\smi(\bm x; C)$, which is commonly used to quantify task-specific information, does not admit an analytical expression for mixture data, we provide analytical expressions for bounds and estimators of $\smi(\bm x; C)$.

\subsection{Problem Statement and Contributions}

We consider the data as a continuous random variable $\bm x$ and the class label as a discrete random variable $C$, where $C$ can be any integer in $[1, \Pi]$ and $\Pi$ is the number of classes. The bold symbol $\bm x$ emphasizes that $\bm x$ is a vector, which can be high-dimensional. We assume that, when restricted to any of the classes, the conditional differential entropy of $\bm x$ is well-defined, or in other words, $(\bm x, C)$ is a good mixed-pair vector \cite{nair2006entropy}. The mutual information between the data $\bm x$ and the class label $C$ can be defined as \cite{beknazaryan2019mutual}
\begin{equation}
\begin{split}
	\smi(\bm x; C) &= \kld{\,\prob(\bm x, C)}{\,\Prob(C)\cdot\prob(\bm x)\,}\\
	& = \sum_{C}\int\mathrm{dx}\,\prob(\bm x, C)\ln{\frac{\prob(\bm x, C)}{\Prob(C)\cdot\prob(\bm x)}}.
\end{split}
\label{eq:MI_definition}
\end{equation}

\indent When $\prob(\bm x)$ is a mixture distribution with $N$ components, 
\begin{equation}
	\prob(\bm x) = \sum_{i=1}^{N} w_{i} \,\prob_i(\bm x),
\end{equation}
where $w_i$ is the weight of component $i$ ($w_i\geq0$ and $\sum_i w_i = 1$), and $\prob_i$ is the probability density of component $i$. The conditional distribution of the data, when the class label is $c$, also follows a mixture distribution.

\indent In this work, we propose new bounds and estimators of the Shannon mutual information between a mixture distribution $\bm x$ and its class label $C$. We provide a lower bound, a variational upper bound and three estimators of $\smi(\bm x; C)$, all based on pair-wise distances. We present closed-form expressions for the bounds and estimators. Furthermore, we use numerical simulations to compare the bounds and estimators to Monte Carlo (MC) simulations and a set of bounds derived from entropy bounds.

\subsection{Related works}

\indent Although estimation of conditional entropy and mutual information has been extensively studied \cite{kozachenko1987sample, ahmad1976nonparametric, laurent1996efficient, basharin1959statistical}, research has focused on purely discrete or continuous data. Nair et al.\ \cite{nair2006entropy} extended the definition of the joint entropy to mixed-pairs, which consists of one discrete variable and one continuous variable. Ross \cite{ross2014mutual}, Moon et al.\ \cite{moon2017ensemble} and Beknazaryan et al.\ \cite{beknazaryan2019mutual} provided methods for estimating mutual information from samples of mixed-pairs based on nearest-neighbor or kernel estimator. Gao et al.\ \cite{gao2017estimating} extended the definition of mutual information to the case that each random variable can have both discrete and continuous components through the Radon-Nikodym derivative. Here our goal is to study mutual information for mixed-pairs, where the data $\bm x$ is continuous and the class label $C$ is discrete.

\indent When the underlying distribution of the data is unknown, the mutual information can be approximated from samples with a number of density or likelihood-ratio estimators based on binning \cite{darbellay1999estimation, moddemeijer1989estimation}, kernel methods\cite{fraser1986independent, moon1995estimation, kandasamy2015nonparametric}, k-nearest-neighbor (kNN) distances \cite{kraskov2004estimating, singh2016finite}, or approximated Gaussianity (Edgeworth expansion \cite{hulle2005edgeworth}). To accommodate high dimensional data (such as image and text) or large datasets, Gao et al.\ \cite{gao2015efficient} improved the kNN estimator with a local non-uniformity correction term; Jiao et al.\ \cite{jiao2015minimax} proposed a minimax estimator of entropy that achieves the optimal sample complexity; Belghazi et al.\ \cite{belghazi2018mutual} presented a general purpose neural-network estimator; Poole et al.\ \cite{poole2019variational} provided a thorough review and several new bounds on mutual information that is capable to trade off bias for variance.

\indent  However, when the underlying data distribution is known, the exact computation of mutual information is tractable only for a limited family of distributions \cite{michalowicz2013handbook, nielsen2010entropies}. The mutual information for mixture distributions has no known closed-from expression \cite{carreira2000mode, michalowicz2008calculation, zobay2014variational}; hence MC sampling and numerical integration are often employed as unbiased estimators. MC sampling of sufficient accuracy is computationally intensive \cite{chen2008accelerated}. Numerical integration is limited to low-dimensional problems \cite{joe1989estimation}. To reduce the computational requirement, deterministic approximations have been developed using merged Gaussian \cite{hershey2007approximating, nielsen2017maxent}, component-wise Taylor-series expansion \cite{huber2008entropy}, unscented transform \cite{julier1996general} and pair-wise KL divergence between matched components \cite{goldberger2003efficient}. The merged Gaussian and unscented transform estimators are biased, while the Taylor expansion method provides a trade-off between computational demands and accuracy. 

\indent Two papers that have deeply inspired our work are \cite{hershey2007approximating} and \cite{kolchinsky2017estimating}. Hersey et al.\ \cite{hershey2007approximating} proposed a variational upper bound and an estimator of the KL divergence between two Gaussian mixtures by pair-wise KL divergence. Hersey et al.\ \cite{hershey2007approximating} has shown empirically that the variational upper bound and estimator perform better than other deterministic approximations, such as merged Gaussian, unscented transform and matched components. Kolchinsky et al. \cite{kolchinsky2017estimating} has bounded entropy of mixture distributions with pair-wise KL divergence and Chernoff-$\alpha$ ($C_\alpha$) divergence and demonstrated through numerical simulations that these bounds are tighter than other well-known existing bounds, such as the kernel-density estimator \cite{joe1989estimation, hall1993estimation} and the expected-likelihood-kernel estimator \cite{jebara2003bhattacharyya, jebara2004probability, contreras2016bounds}. Our results are not obvious from either paper, as the calculation of $\smi(\bm x; C)$ involves a summation of multiple entropies or KL divergences. Instead of providing bounds for each term (entropy) in the summation, we directly bound and estimate the mutual information. 

\section{Main Results} \label{sec:results}
\indent In this section, we provide three estimators of $\smi(\bm x; C)$ and a pair of lower and upper bounds. All bounds and estimators are based on pair-wise KL divergence and $C_\alpha$ divergence. Furthermore, we provide proofs of the lower and upper bounds. Before presenting our main results, we start with a few definitions. The marginal distribution on the class label $C$ is 
\begin{equation}
	\Prob(C=c)  = P_c = \sum_{i\in\{c\}} w_i.
\end{equation}
Note that $\sum_{c=1}^{\Pi} P_c = 1$ and $\{c\}$ is the set of the components that have class label $C=c$. The conditional distribution of the data, when the class label is $c$, is given by
\begin{equation}
\prob(\bm x|c) = \sum_{i\in\{c\}}\frac{ w_i}{P_c}\,\prob_i(\bm x).
\end{equation}
Expressing the marginal distribution in terms of the conditional distribution, we have
\begin{equation}
	\prob(\bm x) = \sum_c P_c\cdot\prob(\bm x|c).
\end{equation}
The joint distribution of the data and class label is
\begin{equation}
	\prob(\bm x, c) = P_c\cdot\prob(\bm x|c) = \sum_{i\in\{c\}} w_i\prob_i(\bm x)
\end{equation}

%

\subsection{Pair-wise distances}
\indent The Kullback-Leibler (KL) divergence is defined as
\begin{equation}
 \kld{\prob_i}{\prob_j} = \int\mathrm{d\bm x}~\prob_i(\bm x)\ln\frac{\prob_i(\bm x)}{\prob_j(\bm x)}.
\end{equation}

\indent The $C_\alpha$ divergence \cite{chernoff1952measure} between the two distribution $\prob_i(\bm x)$ and $\prob_j(\bm x)$ is defined as
\begin{equation}
 	\ca{\prob_i}{\prob_j}  = -\ln\int\mathrm{d\bm x}\,\prob_i^\alpha(\bm x) \,\prob_j^{1-\alpha}(\bm x),
\end{equation}
for real-valued $\alpha\in[0,1]$. More specifically, when $\alpha=1/2$, the Chernoff divergence is Bhattacharyaa distance.\\

\subsection{Bounds and estimates of the mutual information}

\indent We adopt the convention that $\ln0=0$ and $\ln(0/0)=0$. An exact expression of the mutual information is
\begin{equation}
\smi(\bm x;C) = \ent(C) - \sum_{i=1}^N w_i \mathrm{E}_{\prob_i} \left[\ln{\frac{\sum_{j=1}^N w_j\prob_j}{\sum_{k\in \{C_i\}} w_k\prob_k}}\right],
\label{eq:I_exact}
\end{equation}
where $\{C_i\}$ is the set of component index that is in the same class with component $i$ and $\mathrm{E}_{\prob_i}[f]=\int\mathrm{dx}\,\prob_i(\bm x) f(\bm x)$ is the expectation of $f$ with respect to the probability density function $\prob_i$.

\indent Two approximations of $\smi(\bm x; C)$ are  
\begin{equation}
\begin{split}
	\mathrm{{\hat I}_{C_\alpha}}(\bm x;C) &= \ent(C) - \sum_{i=1}^N w_i  \ln{\frac{\sum_{j=1}^N w_j e^{-\ca{\prob_i}{\prob_j}}}{\sum_{k\in \{C_i\}} w_k e^{-\ca{\prob_i}{\prob_k}}}}, \\
	\mathrm{{\hat I}_{KL}}(\bm x;C) &= \ent(C) - \sum_{i=1}^N w_i \ln{\frac{\sum_{j=1}^N w_j e^{ -\kld{\prob_i}{\prob_j}}}{\sum_{k\in \{C_i\}} w_k e^{ -\kld{\prob_i}{\prob_k}}}}.
\end{split}
\label{eq:I_estimates}
\end{equation}

\indent Another approximation of $\smi(\bm x; C)$ is 
\begin{equation}
\begin{split}
	&\mathrm{{\hat I}_{KL\&C_\alpha}}(\bm x;C) = \ent(C) - \sum_{i=1}^N w_i  \ln{\frac{\sum_{j=1}^N w_j e^{-D_{ij}}}{\sum_{k\in \{C_i\}} w_k e^{-D_{ik}}}}, \\
	&\text{where}\quad \frac{1}{D_{ij}} = \frac{1}{2}\left(\frac{1}{\kld{\prob_i}{\prob_j}} + \frac{1}{{\ca{\prob_i}{\prob_j}}}\right).
\end{split}
\label{eq:I_estimates}
\end{equation}
As $D$ is a function of both KL and $C_\alpha$ divergences, we denote this estimator with the subscript `$KL\&C_\alpha$'. 

\indent A lower bound on $\smi(\bm x; C)$ based on pair-wise $C_\alpha$ is
 \begin{equation}
 	\begin{split}
	&\mathrm{{ I}_{lb\_{C_\alpha}}} = -\sum_{c=1}^{\Pi} P_c\ln\left[\sum_{c'=1}^{\Pi}P_{c'} \cdot\text{min}(1,Q_{cc'})\right],\,\,where\\
	&Q_{cc'} =\sum_{i\in \{c\}}\sum_{j\in \{c'\}} \left(\frac{w_i}{P_c}\right)^{\alpha_c}\left(\frac{w_j}{P_{c'}}\right)^{1-\alpha_c} e^{-C_{\alpha_c}(\prob_i||\prob_j)},\\
	\end{split}
	\label{eq:I_cCalpha}
\end{equation}
and min$(\cdot)$ is the minimum value function. 

\indent A variational upper bound on $\smi(\bm x, C)$ based on pair-wise KL is
\begin{equation}
\mathrm{{ I}_{ub\_KL}} = \ent(C) - \sum_m\sum_{c=1}^{\Pi} \phi_{m,c}\ln\frac{\sum_{c'=1}^{\Pi}\phi_{m,c'} e^{-KL(\prob_{m,c}, \prob_{m,c'})}}{\phi_{m,c}},
\label{eq:I_bKL}  
\end{equation}
where $\phi_{m,c}$ are the variational parameters and detailed explanation of the notations is given in Section~\ref{ssec:proof_variation_ub}.

%
 \subsection{Proof of the lower bound}
\indent We propose a lower bound on $\smi(\bm x; C)$ based on pair-wise $C_\alpha$ divergences. For ease of notation, we denote the conditional distribution $\prob(\bm x|C=c)$ as $\prob_c$. We first make use of a derivation from \cite{kolchinsky2017estimating} and \cite{haussler1997mutual} to bound $\smi(\bm x; C)$ with class-wise divergence $C_\alpha(\prob_c, \prob_{c'})$:
 \begin{equation}
 	\begin{split}
	\smi(\bm x; C)  =& \sum_{c}P_c\int\mathrm{dx}\,\prob_c\cdot\ln{\frac{\prob_c}{\prob(\bm x)}} \\
	=&- \sum_{c}P_c\int\mathrm{dx}\,\prob_c\cdot\ln{\frac{\sum_{c'}P_{c'}\prob_{c'}^{1-\alpha_c}}{\prob_c^{1-\alpha_c}}}\\ 
	&-\sum_{c}P_c\int\mathrm{dx}\,\prob_c\cdot\ln\frac{\prob(\bm x)}{\prob_c^{\alpha_c}\sum_{c'}P_{c'}\prob_{c'}^{1-\alpha_c}} \\
	\geq& - \sum_{c}P_c\ln{\sum_{c'}P_{c'}\int\mathrm{dx}\,\prob_c^{\alpha_c}\cdot\prob_{c'}^{1-\alpha_c}}\\ 
	&-\ln\int\mathrm{dx}\,\sum_{c}P_c\prob_c^{1-\alpha_c}\cdot\frac{\prob(\bm x)}{\sum_{c'}P_{c'}\prob_{c'}^{1-\alpha_c}} \\\\
	=& -\sum_c P_c\ln\left[\sum_{c'}P_{c'} e^{-C_{\alpha_c}(\prob_c||\prob_{c'})}\right]\\
	\end{split}
\end{equation}
This inequality follows from Jensen's inequality and the convexity of function $\ln(x)$. The parameter $\alpha_c$, which is specific for a class $c$, can be any value in $[0,1]$. 

\indent The class-wise $C_\alpha$ divergence has a minimum value of zero and the minimum is achieved when the two class has the same distribution. Furthermore, we can bound the $C_\alpha$ divergence through the subadditivity of the function $f(x)=x^\alpha$ when $0\leq\alpha\leq1$. In other words, as $f(a+b)\leq f(a) + f(b)$ for $a\geq0$ and $b\geq0$, the $C_\alpha$ divergence between the conditional distributions $\prob_c$ and $\prob_{c'}$ can be bounded by:
\begin{equation}
\begin{split}
e&^{-C_\alpha(\prob_c||\prob_{c'})} 
= \int\mathrm{dx}\, \left[\sum_{i\in\{c\}} \frac{w_i}{P_c}\prob_i\right]^\alpha\left[\sum_{j\in \{c'\}}\frac{w_j}{P_{c'}}\prob_j\right]^{1-\alpha}\\
&\leq \text{min}\left[1,\sum_{i\in \{c\}}\sum_{j\in \{c'\}} \left(\frac{w_i}{P_c}\right)^\alpha\left(\frac{w_j}{P_{c'}}\right)^{1-\alpha} e^{-C_\alpha(\prob_i||\prob_j)}\right],\\
&= \text{min} (1, Q_{cc'}).\\
\end{split}
\end{equation}
Therefore, Equation~\ref{eq:I_cCalpha} is a lower bound on $\smi(\bm x; C)$.

\indent The best possible lower bound can be obtained by finding the parameters $\alpha_c$ that maximize $\mathrm{{ I}_{lb\_{C_\alpha}}}$, which is equivalent to minimize $\sum_{c'} P_{c'}\text{min} (1, Q_{cc'})$. In a special case when all components are symmetric and identical except the center location, eg. homoscedastic Gaussian mixture, $e^{-C_\alpha(\prob_i||\prob_j)}$ achieves minimum value at $\alpha=1/2$ \cite{kolchinsky2017estimating}. 

\subsection{Proof of the variational upper bound} \label{ssec:proof_variation_ub}
\indent Here we propose a direct upper bound on the mutual information $\smi(\bm x; C)$ using a variational approach. The underlying idea is to match components from different classes. To pick one component from each class, there are $N_1\times N_2...\times N_{\Pi}$ combinations, where $N_c$ is the number of components in class $c$. Denote an integer $M=\prod_{c=1}^{\Pi} N_c$. A component $i$ in class $c$ can be split into $M/N_c$ components with each component corresponding to a component-combination in the other $\Pi-1$ classes. Mathematically speaking, we introduce the variational parameters $\phi_{ij}\geq0$ satisfying the constraints $\sum_{j=1}^{M/N_c} \phi_{ij} = w_i$. Using the variational parameters, we can write the joint distribution as
\begin{equation}
	\begin{split}
		\prob(\bm x, c) = \sum_{i\in\{c\}} w_i\prob_i = \sum_{i\in\{c\}}\sum_{j=1}^{M/N_c} \phi_{ij}\prob_i.
	\end{split}
\end{equation}
Note that the set $\{c\}$ has $N_c$ components. By rearranging indices $(i,j)$ into a vector $m$ of length $M$, we can simplify the joint distribution to
\begin{equation}
		\prob(\bm x, c) = \sum_{m=1}^M \phi_{m,c} \prob_{m,c}(\bm x),
\end{equation}
where the subscript $c$ emphasizes that each class has a unique mapping from $(i,j)$ to $m$ and $\prob_{m,c}(\bm x)$ equals to the corresponding $\prob_i(\bm x)$. 

\indent With this notation, the marginal distribution of the data $\bm x$ is 
\begin{equation}
		\prob(\bm x) = \sum_{c=1}^{\Pi} \prob(\bm x, c)= \sum_{c=1}^{\Pi}\sum_{m=1}^M \phi_{m,c} \prob_{m,c}(\bm x).
\end{equation}
We further define a mini-batch $m$ as
\begin{equation}
   b_m(\bm x)= \sum_{c=1}^{\Pi} \phi_{m,c} \prob_{m,c}(\bm x).
\end{equation}
Each mini-batch contains $\Pi$ components with one component from each class. With this definition, the marginal distribution of the data can be written as $\prob(\bm x) = \sum_{m=1}^M b_m(\bm x)$. The probability of a component in the $m^{th}$ batch is $P_m = \sum_{c} \phi_{m,c}$. The probability density function of the $m^{th}$ batch is $\prob(\bm x|m) = b_m(\bm x)/P_m$.  

\indent  Now we use Jensen's inequality, or more specificly log-sum inequality \cite{cover2012elements}, to bound $\smi(\bm x; C)$ by batch-conditional entropy, 
\begin{equation}
\begin{split}
&\smi(\bm x; C) = \ent(C) + \sum_{c}\int\mathrm{dx}\,\prob(\bm x,c)\cdot\ln{\frac{\prob(\bm x,c)}{\prob(\bm x)}} \\
&= \ent(C) + \sum_c \int\mathrm{dx} \left(\sum_{m} \phi_{m,c}\prob_{m,c}\right)\,\ln\frac{\sum_{m} \phi_{m,c}\prob_{m,c}}{\sum_{m} b_m}  \\
&\leq \ent(C) + \sum_c \int\mathrm{dx} \sum_{m} \left( \phi_{m,c}\prob_{m,c}\ln\frac{\phi_{m,c}\prob_{m,c}}{b_m}\right) \\
	&=\ent(C) + \ent(\bm x|m) -  \ent(C|m)- \sum_{i=1}^{N} w_i\ent_i(\bm x) \\
\end{split}
\end{equation}
where $\ent(C) = -\sum_c P_c \ln P_c$ is the entropy of the class label;  $\ent(\bm x|m)=\sum_m P_m \ent(\prob(\bm x|m))$ is the batch-conditional entropy of the data; $\ent(C|m) = \sum_{m} P_m\ent_m(C)$ is the batch-conditional entropy of the label, where $\ent_m(C)$ is the entropy of the class label for batch $m$, and $\ent_i(\bm x) = \ent(\prob_i(\bm x))$ is the entropy of the $i^{th}$ component. 

\indent We can further bound the batch-conditional entropy with pair-wise KL divergence as 
\begin{equation}
\begin{split}
&\smi(\bm x; C) \leq \ent(C) +\hat\ent_{KL}(\bm x|m) - \ent(C|m) - \sum_{i=1}^{N} w_i\ent_i(\bm x) \\
          & = \ent(C) - \sum_m\sum_c \phi_{m,c}\ln\frac{\sum_{c'}\phi_{m,c'} e^{-KL(\prob_{m,c}, \prob_{m,c'})}}{\phi_{m,c}}\\
          &:=\mathrm{{ I}_{ub\_{KL}}}
\end{split}
\end{equation}
where $\hat\ent_{KL}(\bm x|m)$ is an upper bound of the batch-conditional entropy and the inequality has been proved in \cite{kolchinsky2017estimating}. 

\indent The tightest upper bound attainable through this method can be found by varying parameters $\phi_{m,c}$ to minimize $\mathrm{{ I}_{ub\_{KL}}}$. The minimization problem has been proved to be convex (see Appendix~A
). The upper bound $\mathrm{{ I}_{ub\_{KL}}}$ can be minimized iteratively by fixing the parameters $\phi_{m,c'}$ (where $c'\neq c$) and optimizing parameters $\phi_{m,c}$ under linear constraints. At each iteration step $\mathrm{{ I}_{ub\_{KL}}}$ is lowered, and the convergent is the tightest variational upper bound on the mutual information.

\indent Non-optimum variational parameters still provide upper bounds on $\smi(\bm x; C)$. There are $M\times\Pi$ variational parameters, $M\times\Pi$ non-equality constraints and $N$ equality constraints. When the number of classes or components is large, the minimization problem will be computationally intensive. A non-optimum solution that is similar to the matched bound \cite{hershey2007approximating, do2003fast} can be obtained by dividing all components into max($N_c$) mini-batches by matching each component $i$ to one component in each class. Mathematically speaking, $\phi_{ij}=w_i$ for one pair of matched $(i,j)$ and $\phi_{ij}=0$ otherwise. To find the mini-batches, the Hungarian method \cite{munkres1957algorithms, kuhn1955hungarian} for assignment problems can be applied. 
 
\section{Numerical simulations} \label{sec:numerical}

\begin{figure*} [htbp!]
\begin{subfigure}{0.44\textwidth}
\centering
\includegraphics[width=0.95\linewidth]{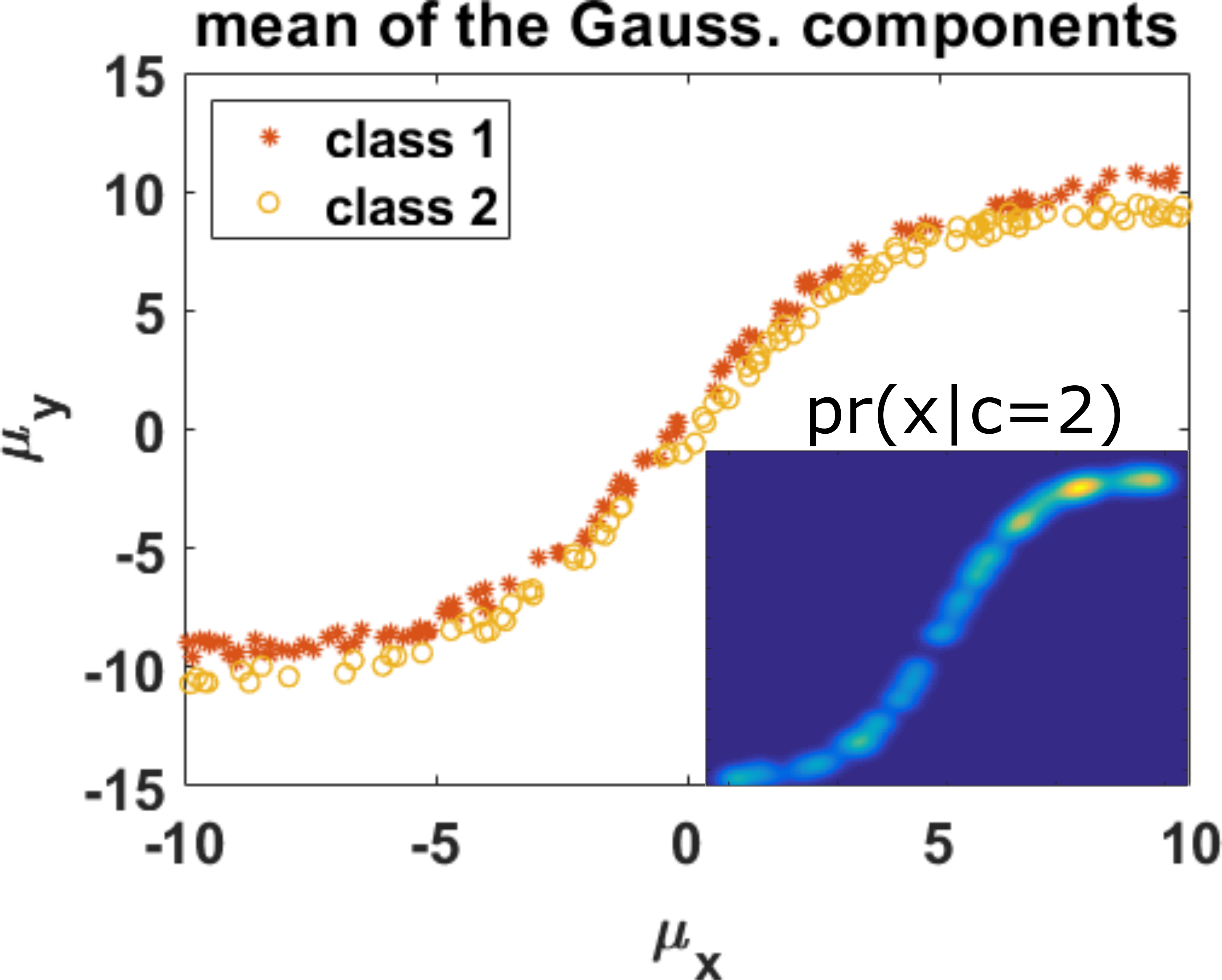}
\caption{}
\label{fig:scene1_pos}
\end{subfigure}
\begin{subfigure}{0.56\textwidth}
\centering
\includegraphics[width=0.95\linewidth]{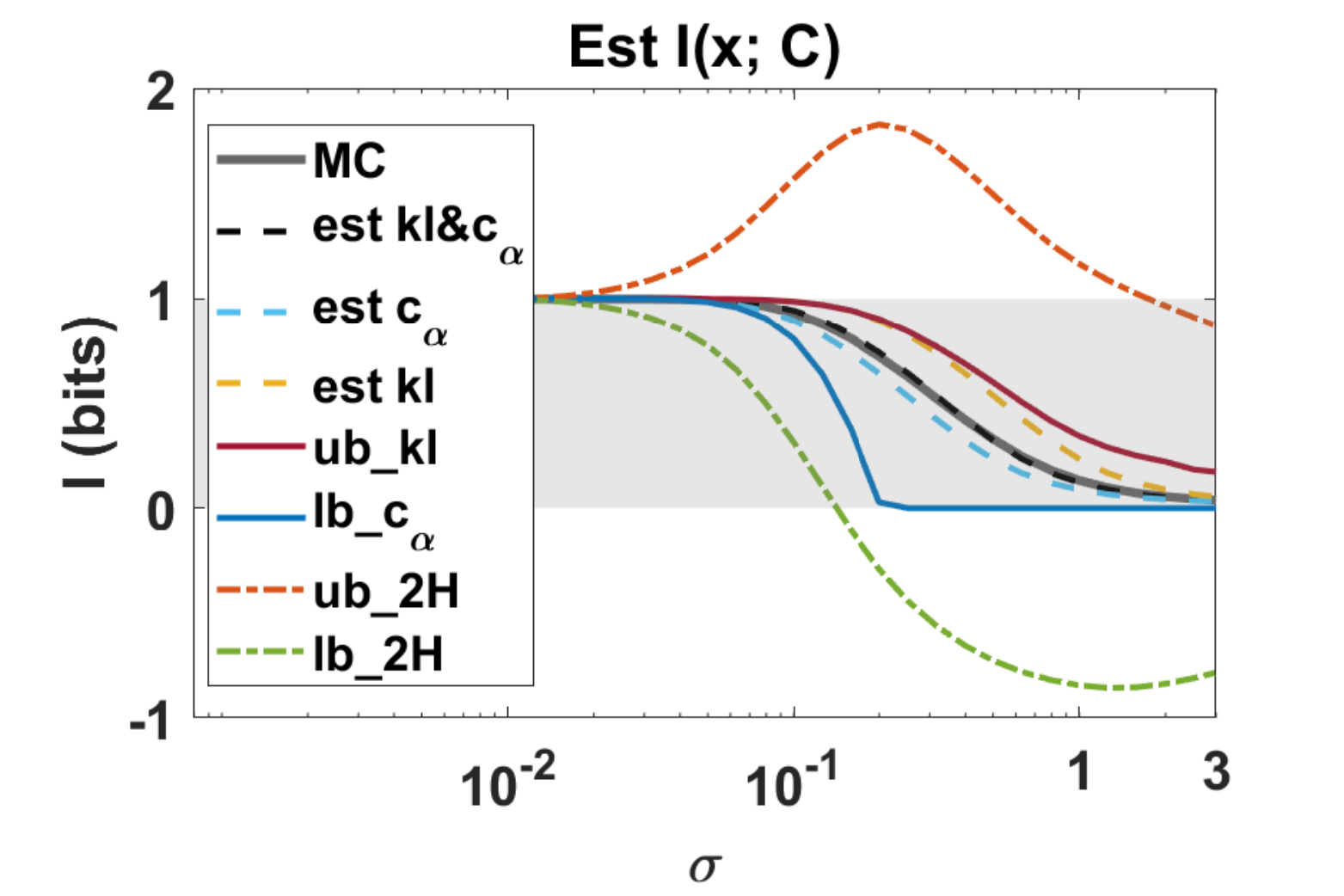}
\caption{}
\label{fig:scene1_smi}
\end{subfigure}
\caption{(a) The locations of the center of the components and the mixture distribution $\prob(\bm x|c=2)$ when $\sigma=0.5$ (insert). (b) Estimates of $\smi(\bm x; C)$.}
\end{figure*}

\indent In this section, we run numerical simulations and compare estimators on mutual information between mixture data and class labels. We consider a simple example of binary classification of mixture data, where the mixture components are two-dimensional homoscedastic Gaussians. The component centers are close to the class boundary and uniformly distributed along the boundary. The location of the component centers are plotted in the Figure~\ref{fig:scene1_pos}, where the component centers are represented by a red star (class 1) or a yellow circle (class 2). Each class consists 100 two-dimensional Gaussian components with equal weights. The components have the same covariance matrix $\sigma^2I$, where $I$ is the identity matrix and $\sigma$ represents the size of the Gaussian components. The conditional distribution $\prob(\bm x|c=2)$ is plotted in the insert of Figure~\ref{fig:scene1_pos} for $\sigma=0.5$. When $\sigma$ is larger, the components of the mixtures distribution are more connected; when $\sigma$ is smaller, the components are more isolated. Estimates of $\smi(\bm x; C)$ are calculated for varying $\sigma$. 

\indent A pair of obvious bounds of $\smi(\bm x; C)$ are $[0, \ent(C)]$, where $\ent(C)$ is the entropy of the class label $\Prob(C)$. Another pair of upper and lower bounds of $\smi(\bm x;C)$ can be derived from bounds on mixture entropy as
\begin{equation}
\begin{split}
 \mathrm{{I}_{lb\_2H}}&=\ent_{lb}(\bm x) - \ent_{ub}(\bm x|C)\\ 
 \mathrm{{I}_{ub\_2H}}&=\ent_{ub}(\bm x) - \ent_{lb}(\bm x|C), 
\end{split}
\label{eq:I_CK_KC}
\end{equation}
where the upper and lower bound of entropy based on pair-wise KL and $C_\alpha$ divergences have been provided by \cite{kolchinsky2017estimating}. These bounds on $\smi(\bm x, C)$ are based on two entropy bounds, hence the subscript `2H'. 

\indent We evaluate the following estimates of $\smi(\bm x, C)$:
\begin{enumerate}
	\item The new variational upper bound and the new lower bound, $\mathrm{{ I}_{ub\_{KL}}}$ and $\mathrm{{ I}_{lb\_{C_\alpha}}}$, are plotted in dark red and blue solid lines, respectively.
	\item The estimates based on the pair-wise KL, $C_\alpha$ or D (a function of both KL and $C_\alpha$ divergences) are plotted in yellow, light blue and black dashed lines, respectively.
	\item The true mutual information, $\smi(\bm x, C)$, as estimated by MC sampling of the mixture model (grey solid line). 
	\item The upper and lower bounds $\mathrm{{I}_{lb\_2H}}$ and $\mathrm{{I}_{ub\_2H}}$ are plotted in orange and green dot-dashed lines, respectively.
\end{enumerate}
The obvious bounds on $\smi(\bm x, C)$, which are $[0,H(C)]$, are also presented by an area in grey. The Monte-Carlo simulation results, which can serve as the benchmark, are calculated with $10^6$ samples. We use $\alpha=1/2$ in the calculation of $C_\alpha$ divergences, as it provides the optimum bounds for our example. We also present details of our implementation and results of two other scenarios in Appendix~B. 

\indent Our new upper bound and lower bound appear to be tighter than the bounds derived from entropy bounds over the range of $\sigma$ considered in our simulation. In Figure~\ref{fig:scene1_smi}, where the estimates of $\smi(\bm x, C)$ are plotted, the results show that the blue and dark red solid lines are almost always within the area covered by the green and orange dot-dashed lines. The three estimates, $\mathrm{{\hat I}_{KL}}$, $\mathrm{{\hat I}_{C_\alpha}}$ and $\mathrm{{\hat I}_{KL\&C_\alpha}}$, all follow the trend of $\smi(\bm x, C)$. More specifically, $\mathrm{{\hat I}_{KL}}$ (yellow dashed line) follows the new variational upper bound $\mathrm{{ I}_{ub\_{KL}}}$ (deep red solid line) closely; $\mathrm{{\hat I}_{C_\alpha}}$ (blue dashed line) is a good estimator of $\smi(\bm x, C)$ (grey solid line); $\mathrm{{\hat I}_{KL\&C_\alpha}}$ (black dashed line) is another good estimator of $\smi(\bm x, C)$, as the black dashed line tracks the grey solid line closely. The differences between the three estimates and the $\smi(\bm x, C)$ calculated from MC simulation are plotted in Appendix~B. 

\section{Conclusion}
\indent We provide closed-form bounds and approximations of mutual information between mixture data and class labels. The closed-form expressions are based on pair-wise distances, which are feasible to compute even for high-dimensional data. Based on numerical results, the new bounds we proposed are tighter than the bounds derived from bounds on entropy and the approximations serve as good surrogates for the true mutual information. 

\appendices
\section{The Minimization of $\mathrm{{ I}_{ub\_KL}}$} \label{sec:app_convexity}
The minimization problem of $\mathrm{{ I}_{ub\_KL}}$ by varying $\phi_{mc}$ is convex, which we prove in this section. The convexity of the minimization problem can be checked through the first and second-order derivatives. For ease of notation, we define $S_{m,c} = \sum_{c'}\phi_{m,c'} e^{-\mathrm{KL}(\prob_{m,c}, \prob_{m,c'})}$ and $E_{m,cc'}=\exp({-\mathrm{KL}(\prob_{m,c}, \prob_{m,c'})})$. The first derivative of $\mathrm{{ I}_{ub\_{KL}}}$ is:
\begin{equation}
	\frac{\partial {I}_{ub\_{KL}}}{\partial \phi_{m,c}} =-\ln\left(\frac{S_{m,c}}{\phi_{m,c}}\right) - \frac{\phi_{m,c}}{S_{m,c}} - \sum_{c'\neq c}\frac{\phi_{m,c'}E_{m,c'c}}{S_{m,c'}} +1
\end{equation}
The second derivative is:
\begin{equation}
\begin{split}
	H_{cc}&=\frac{\partial^2 {I}_{ub\_{KL}}}{(\partial \phi_{m,c})^2} \\
	&=\frac{(S_{m,c}-\phi_{m,c})^2}{(S_{m,c})^2\phi_{m,c}}+\sum_{c'\neq c}\frac{\phi_{m,c'}(E_{m,c'c})^2}{(S_{m,c'})^2}\\
\end{split}
\end{equation}
for the diagonal terms and
\begin{equation}
\begin{split}
	H_{cc'}&=\frac{\partial^2 {I}_{ub\_{KL}}}{\partial \phi_{m,c} \partial \phi_{m,c'}} \\
	&=\frac{\phi_{m,c}-S_{m,c}}{(S_{m,c})^2}E_{m,cc'} + \frac{\phi_{m,c'}-S_{m,c'}}{(S_{m,c'})^2}E_{m,c'c},\\
	\end{split}
\end{equation}
for $c'\neq c$. For any given vector $\bm\theta$ of length $\Pi$, 
\begin{equation}
\begin{split}
\bm\theta^TH\bm\theta =& \sum_c\left(\theta_c^2H_{cc}+\sum_{c'\neq c}\theta_c\theta_{c'}H_{cc'}\right)\\
=&\sum_c\Bigg[\theta_c^2\frac{(S_{m,c}-\phi_{m,c})^2}{(S_{m,c})^2\phi_{m,c}} +\sum_{c'\neq c}\theta_{c'}^2\frac{\phi_{m,c}(E_{m,cc'})^2}{(S_{m,c})^2} \\
&\quad\quad+\sum_{c'\neq c}2\theta_c\theta_{c'}\frac{\phi_{m,c}-S_{m,c}}{(S_{m,c})^2}E_{m,cc'}\Bigg]\\
=&\sum_c \left[\frac{(S_{m,c}-\phi_{m,c})\theta_c}{S_{m,c}\sqrt{\phi_{m,c}}}-\sum_{c'\neq c}\frac{\sqrt{\phi_{m,c}}E_{m,cc'}\theta_{c'}}{S_{m,c}}\right]^2 \\
\geq&0. 
\end{split}
\end{equation}
Therefore, $\mathrm{{ I}_{ub\_{KL}}}$ is convex when $\phi_{m,c}$ are considered as the variables. 

\section{On the Numerical simulations} \label{sec:app_numexp}
\indent This appendix provides more simulation results and the detailed expressions used in the numerical simulations. The additional results consider two different distributions of the component-center locations. We further present the difference between the the estimated and the true mutual information. Last but not least, the closed form expressions include the KL and $C_\alpha$ divergences between Gaussian components, the bounds on $\smi(\bm x; C)$ derived from entropy bounds, and the relation between Shannon mutual information and bounds on binary classification error ($\pe$). 

\subsection{Numerical simulation results}
\begin{figure*} [htbp!]
\begin{subfigure}{0.44\textwidth}
\centering
\includegraphics[width=0.9\linewidth]{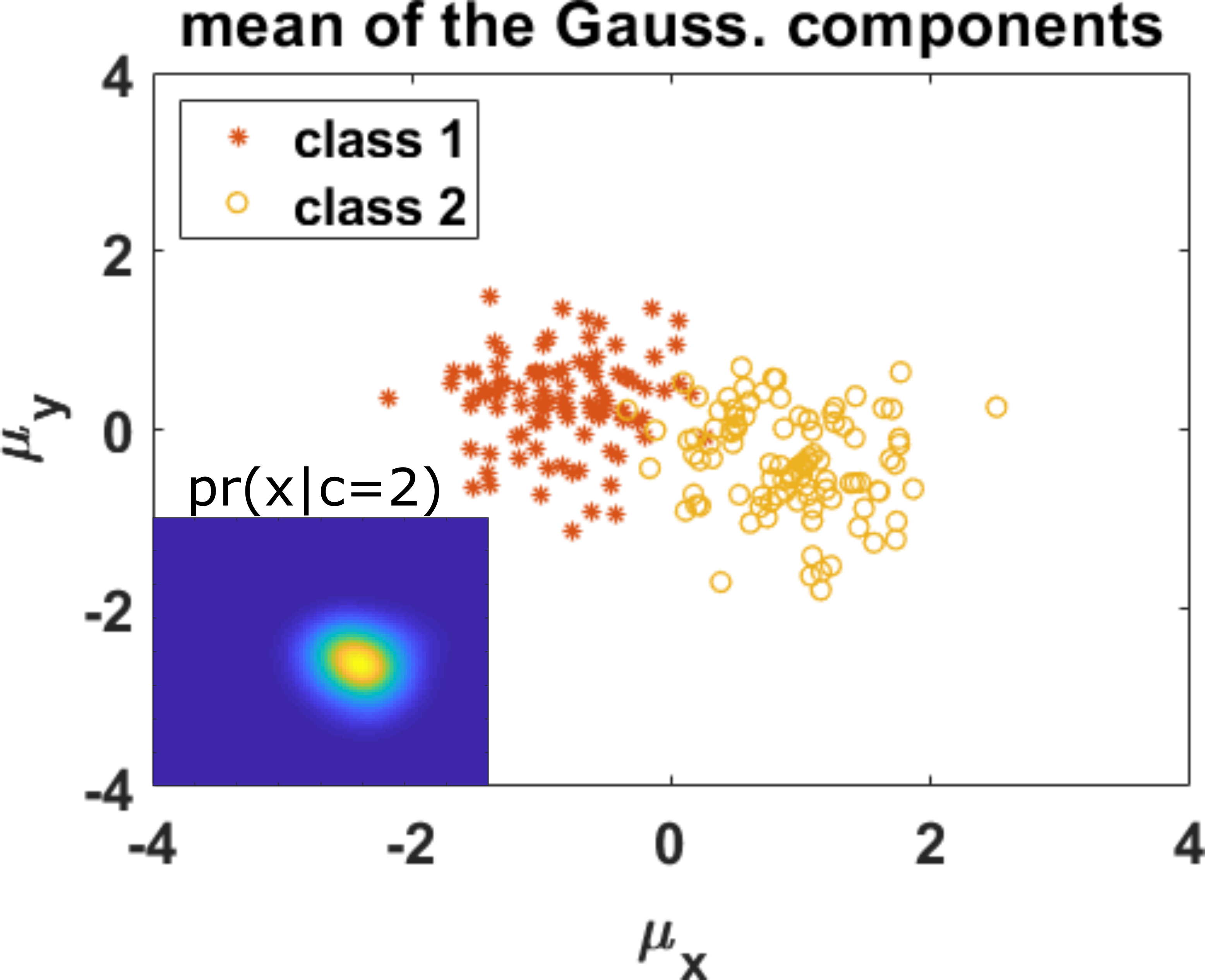}
\caption{}
\label{fig:scene2_pos}
\end{subfigure}
\begin{subfigure}{0.56\textwidth}
\centering
\includegraphics[width=0.9\linewidth]{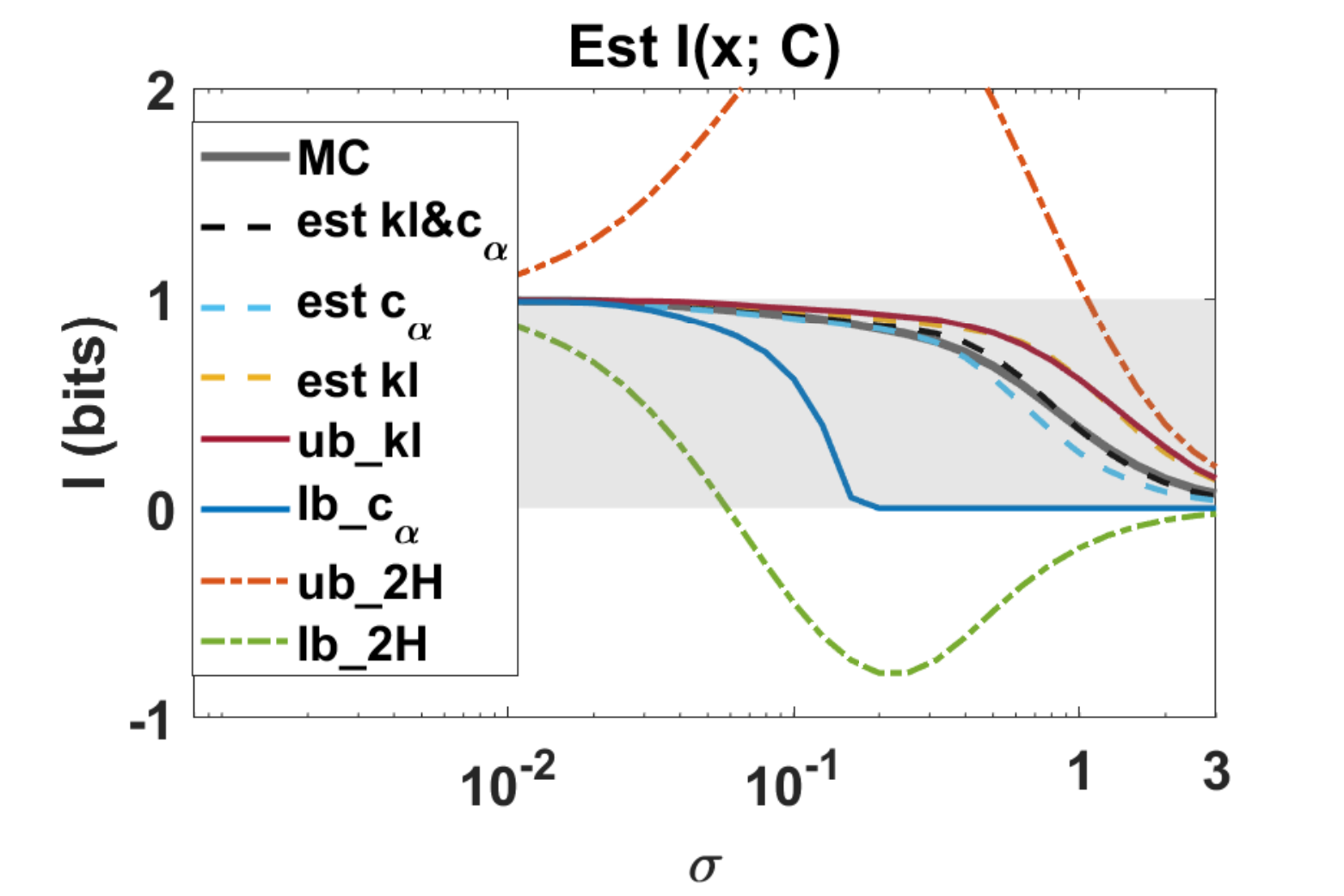}
\caption{}
\label{fig:scene2_smi}
\end{subfigure}

\caption{Scenario 2, where the center of the components are bunched into one group, illustration (a), the mixture distribution $\prob(\bm x|c=2)$ when $\sigma=0.5$ (insert), and estimates of $\smi(\bm x; C)$ (b).}
\end{figure*}

\begin{figure*} [htbp!]
\begin{subfigure}{0.44\textwidth}
\centering
\includegraphics[width=0.9\linewidth]{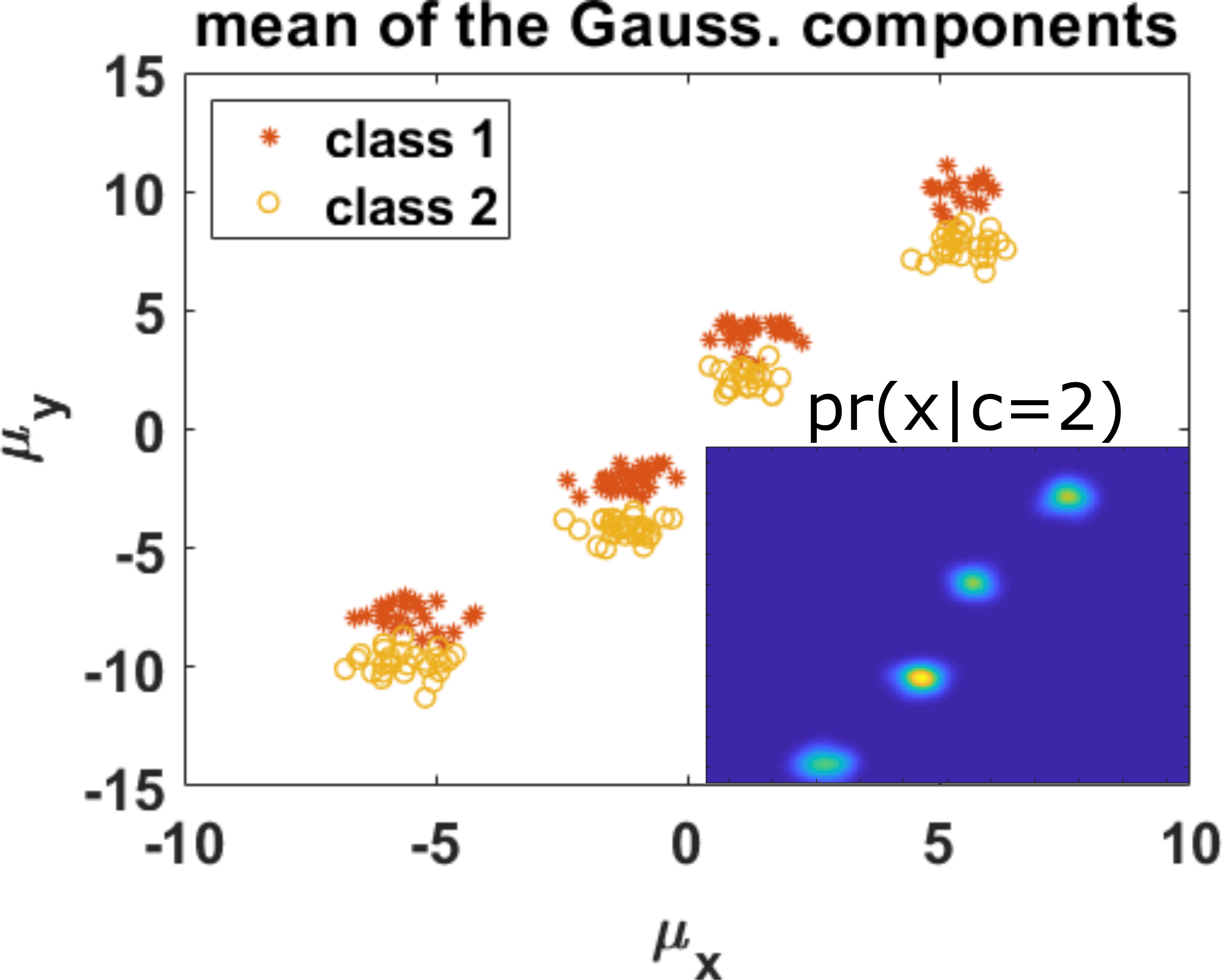}
\caption{}
\label{fig:scene3_pos}
\end{subfigure}
\begin{subfigure}{0.56\textwidth}
\centering
\includegraphics[width=0.9\linewidth]{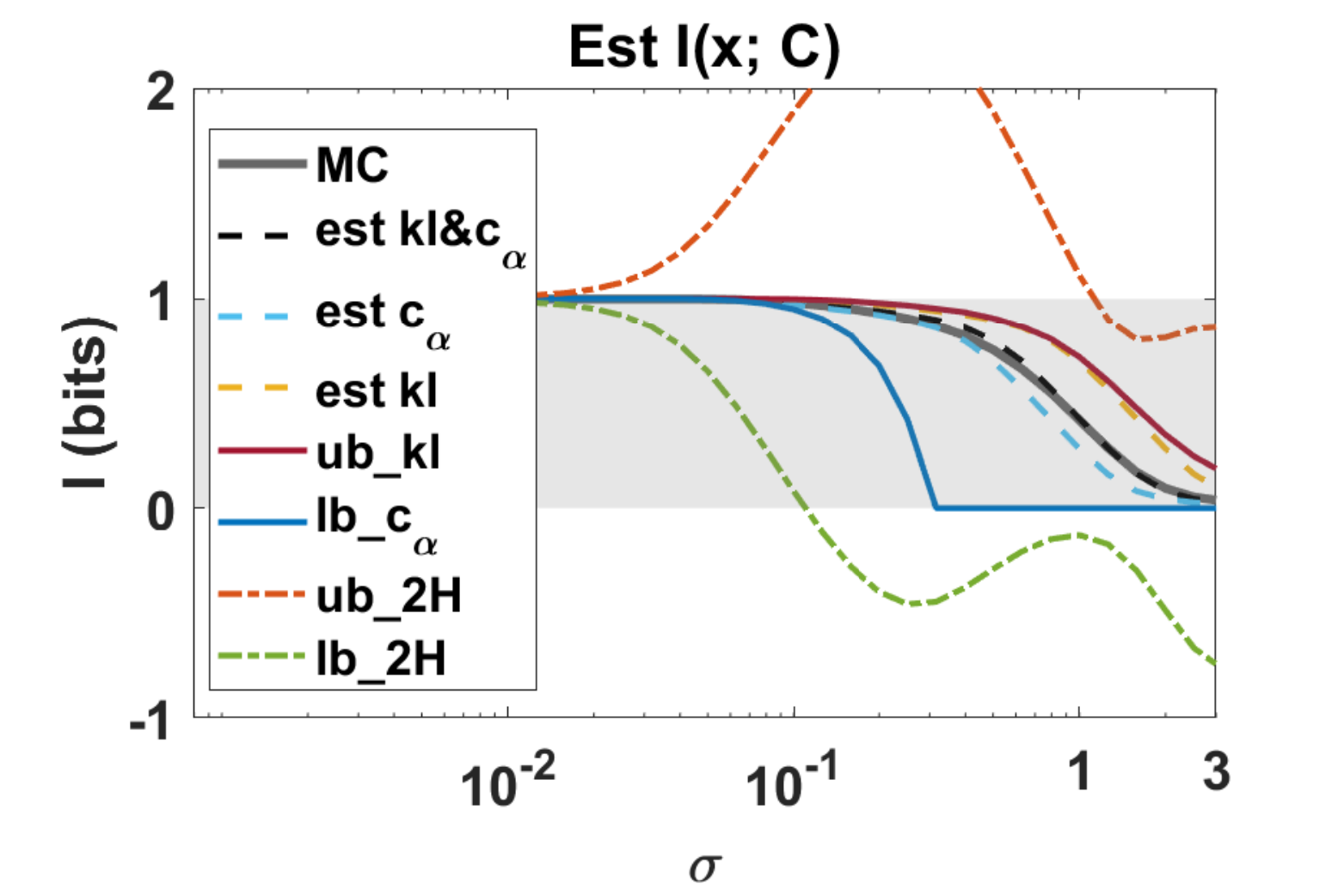}
\caption{}
\label{fig:scene3_smi}
\end{subfigure}
\caption{Scenario 3, where the center of the components are bunched into multiple groups, illustration (a), the mixture distribution $\prob(\bm x|c=2)$ when $\sigma=0.5$ (insert), and estimators of $\smi(\bm x; C)$ (b).}
\end{figure*}

We consider three scenarios: (1) the component centers are uniformly distributed along the class boundary, (2) the component centers are bunched into one group, and (3) the component centers are bunched into several groups. Results on the first scenario has been presented in the Section~\ref{sec:numerical}. In this section, we report on Scenarios 2 and 3. Illustration of the two scenarios are shown in Figure~\ref{fig:scene2_pos} and \ref{fig:scene3_pos}, respectively. 

\begin{figure*} [htbp!]
\begin{subfigure}{0.33\textwidth}
\centering
\includegraphics[width=1\linewidth]{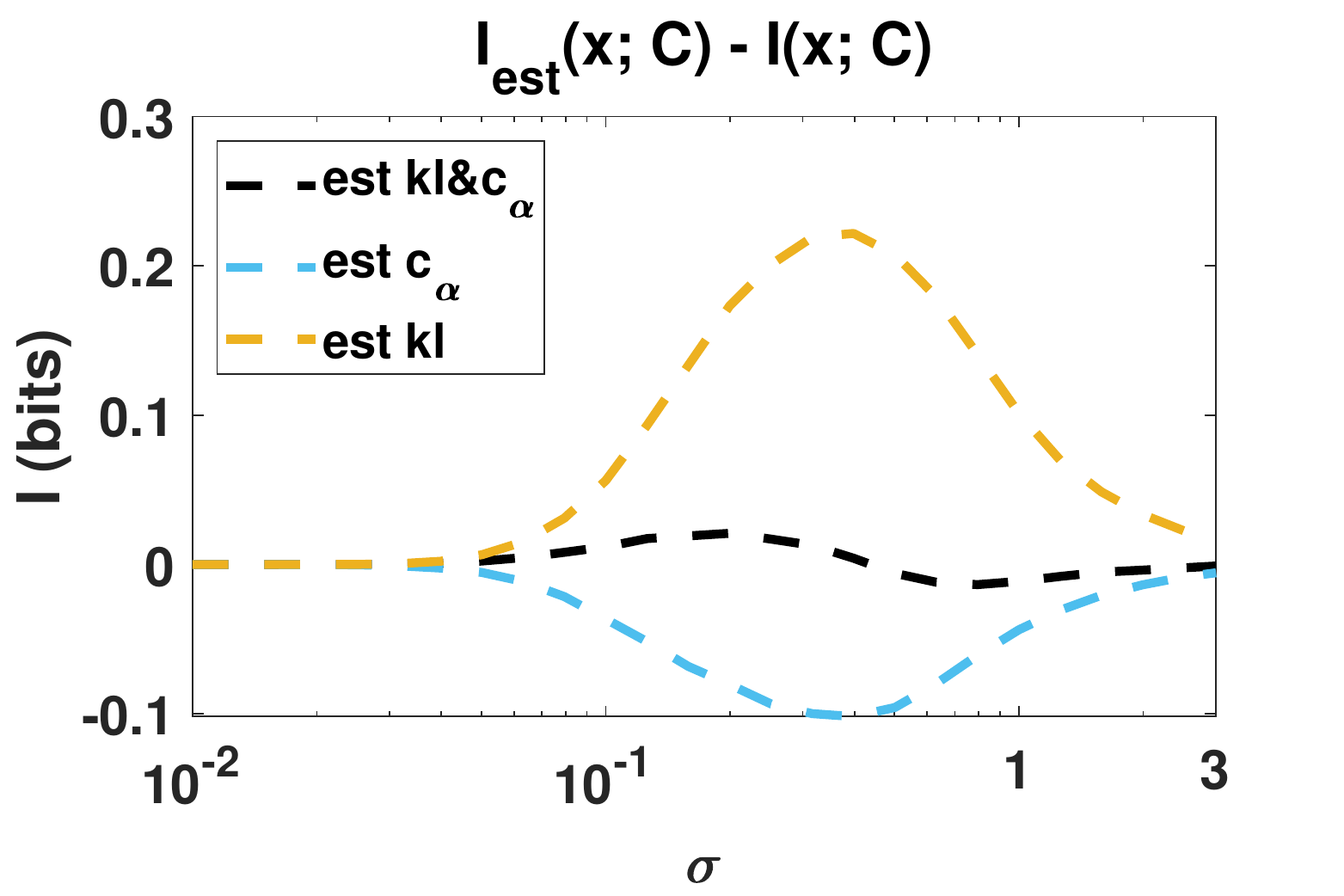}
\caption{}
\end{subfigure}
\begin{subfigure}{0.33\textwidth}
\centering
\includegraphics[width=1\linewidth]{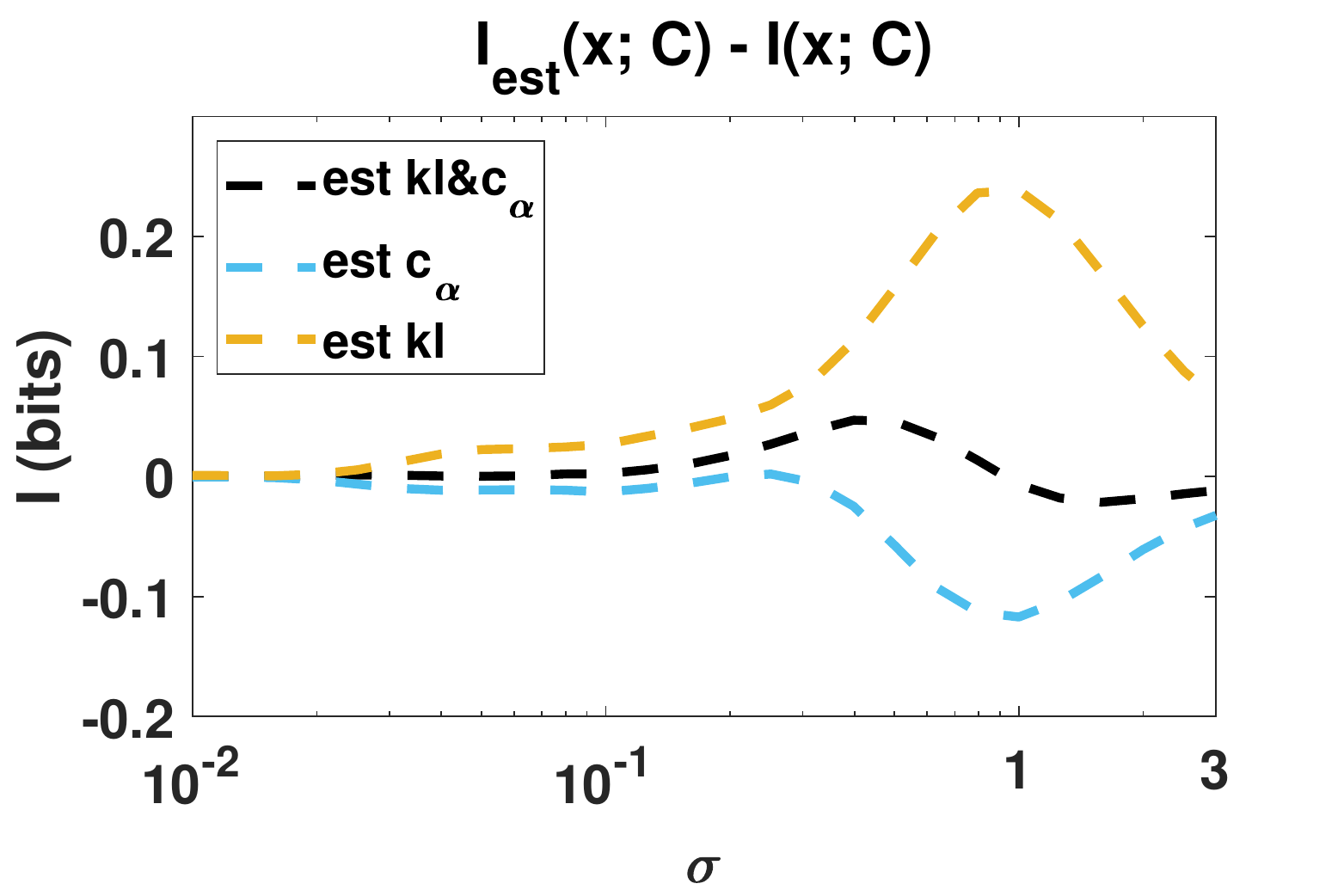}
\caption{}
\end{subfigure}
\begin{subfigure}{0.33\textwidth}
\centering
\includegraphics[width=1\linewidth]{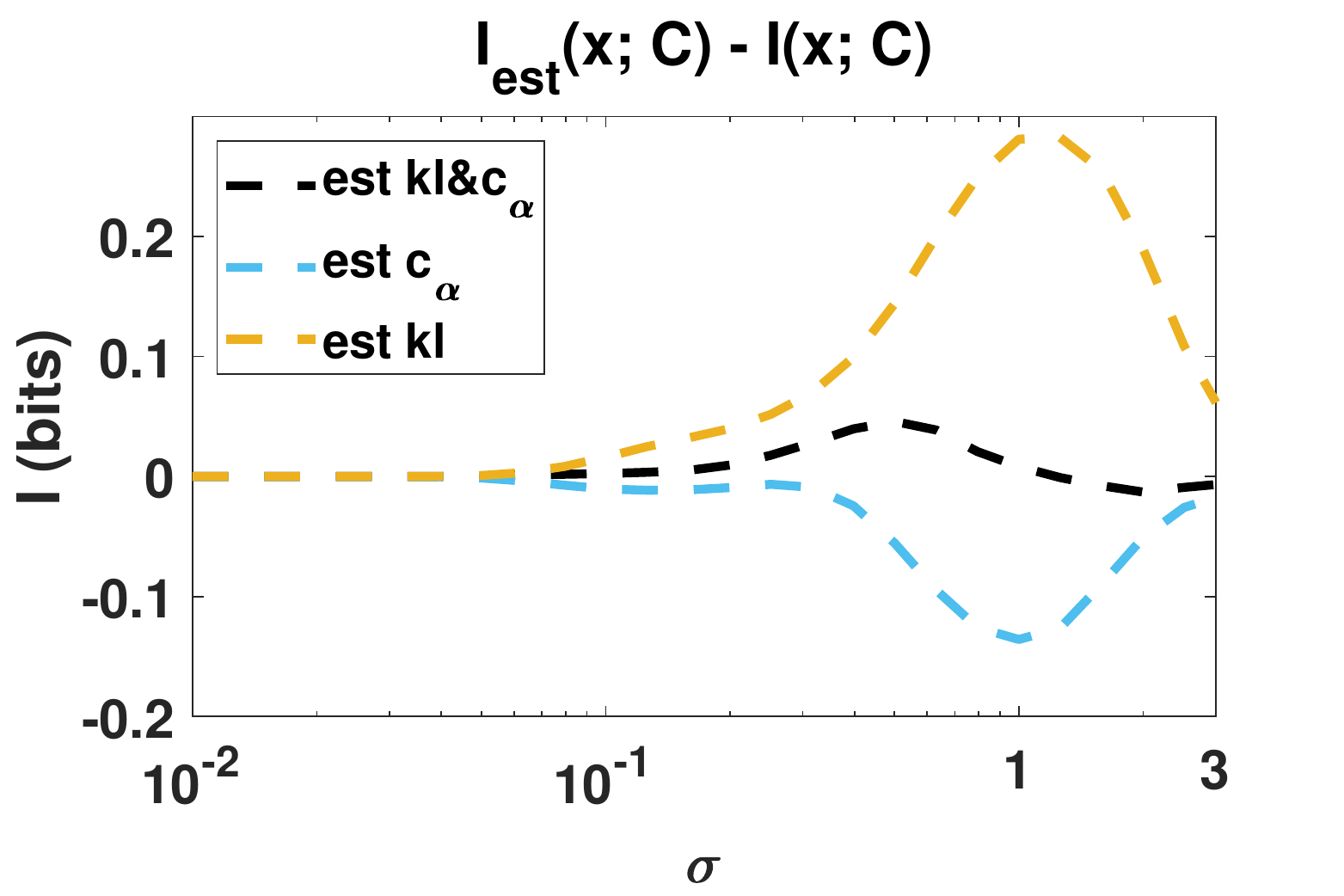}
\caption{}
\end{subfigure}
\caption{$\mathrm{\hat I(\bm x; C)-I(\bm x; C)}$ for (a) Scenario 1, (b) Scenario 2 and (c) Scenario 3. The $\mathrm{I(\bm;C)}$ is calculated from MC simulations. }
\label{fig:diff_MI}
\end{figure*}

\indent The results demonstrated in these two scenarios are similar to that of Scenario 1. To further demonstrate that our estimators are good surrogates for the true mutual information, we plot the difference between the three estimates and the true mutual information calculated from MC sampling in Figure~\ref{fig:diff_MI}. 

\subsection{Closed form expressions for Gaussian mixtures}\label{ssec:app_gmm}
\indent Gaussian functions are often used as components in mixture distributions and have closed form expressions for pair-wise KL and $C_\alpha$ divergences. Denoting the difference in the means of two components as $\bm\mu_{ij}=\bm\mu_i-\bm\mu_j$ and $\Sigma_{\alpha,ij} = (1-\alpha)\Sigma_i+\alpha\Sigma_j$, the $C_\alpha$ divergence between two Gaussian components are
\begin{equation}
\begin{split}
	&\ca{\prob_i}{\prob_j} =\frac{\alpha(1-\alpha)}{2}{\bm\mu^T_{ij}}\Sigma_{\alpha,ij}^{-1}\bm\mu_{ij}+\frac{1}{2}\ln\frac{|\Sigma_{\alpha,ij}|}{|\Sigma_i|^{{1-\alpha}}|\Sigma_j|^{{\alpha}}},
	\end{split}
\end{equation}
where $|\cdot|$ is the determinant. The KL divergence between the same two Gaussian components are 
\begin{equation}
	\kld{\prob_i}{\prob_j}= \frac{1}{2}\left[{\bm\mu^T_{ij}}\Sigma_j^{-1}\bm\mu_{ij}+\ln\frac{|\Sigma_j|}{|\Sigma_i|}\right]+\frac{\text{tr}(\Sigma_j^{-1}\Sigma_i)-d}{2},
\end{equation}
where tr$(\cdot)$ is the trace of the matrix in the parenthesis and $d$ is the dimension of the data. 

\indent When all mixture components have equal covariance matrices $\Sigma_i =\Sigma_j=\Sigma$, we can denote $\lambda_{ij} = {\bm\mu^T_{ij}}\Sigma^{-1}\bm\mu_{ij}$ and have 
\begin{equation}
\begin{split}
	\ca{\prob_i}{\prob_j} &=\alpha(1-\alpha)\lambda_{ij}/2,\\
	\kld{\prob_i}{\prob_j} &= \lambda_{ij}/2.
	\end{split}
\end{equation}
With these expressions, the bounds and estimates of $\smi(\bm x; C)$ have simple forms. \\

\subsection{Expressions of $\mathrm{{I}_{lb\_2H}}$ and $\mathrm{{I}_{ub\_2H}}$}
 \indent The detailed expression for the mutual information bounds derived from entropy bounds are:
\begin{equation}
\begin{split}
 \mathrm{{I}_{lb\_2H}}&=\ent(C) - \sum_{i=1}^N w_i \ln{\frac{\sum_{j=1}^N w_j e^{- \ca{\prob_i}{\prob_j}}}{\sum_{k\in \{C_i\}} w_k e^{-\kld{\prob_i}{\prob_k}}}},\\ 
 \mathrm{{I}_{ub\_2H}}&=\ent(C) - \sum_{i=1}^N w_i \ln{\frac{\sum_{j=1}^N w_j e^{- \kld{\prob_i}{\prob_j}}}{\sum_{k\in \{C_i\}} w_k e^{-\ca{\prob_i}{\prob_k}}}}. 
\end{split}
\label{eq:I_CK_KC}
\end{equation}\\

\subsection{Bounds on $\pe$ for binary classification} \label{ssec:app_pe} 
\indent Bounds on $\smi(\bm x; C)$ can be used to calculate bounds on $\pe$. The Fano's inequality \cite{fano1961transmission} provides a lower bound on $\pe$ for binary classification, as following
\begin{equation}
	\pe \geq h_b^{-1}[\ent(C) - \smi(\mathbf x;C)],
\end{equation}
where $h_b(x) = -x\log_2(x) - (1-x)\log_2(1-x)$ is the binary entropy function, $h_b^{-1}(\cdot)$ is the inverse function of $h_b(\cdot)$. More specifically, one can calculate $P_e$ by placing the value $\ent(C) - \smi(\mathbf x;C)$ on the left side of the binary entropy function and solving for $x$. 

\indent A tight upper bound on binary classification error $\pe$ has been reported recently \cite{hu2016optimization},
\begin{equation}
	\pe \leq {\text{min}} \left\{P_{min},\, f^{-1}[\ent(C) - \smi(\mathbf x; C)] \right\}:=\mathrm{\hat P_{e\_ub}},
	\label{eq:bounds_pe_upper}
\end{equation}
where $P_{min}$ is min$\{P_1, P_2\}$, and $f(x)$ is a function defined by
\begin{equation}
	f(x) = -P_{min}\log_2\frac{P_{min}}{x+P_{min}} - x\log_2\frac{x}{x+P_{min}},
\end{equation}
and $f^{-1}(\cdot)$ is the inverse function of $f(\cdot)$. 

\indent When $\pe\ll1$, $-\pe(\log{\pe}-\log{P_{min}})\lesssim\ent(C) - \smi(\mathbf x;C)\lesssim-\pe\log{\pe}$. Therefore, $\pe$ is on the same order of magnitude as $\ent(C) - \smi(\mathbf x;C)$, when $\pe\ll 1$. \\

\subsection{$\pe$ estimates}
When $\sigma$ is small, all estimates of $\smi(\bm x, C)$ converges to 1. To compare the estimates for $0.1>\sigma>0.01$, we calculate and present an estimate of $\pe$ in this section. This estimate of $\pe$ is the upper bound on $\pe$ presented in the previous section. When a lower bound of $\smi(\bm x, C)$ is used in the calculation (blue solid line and green dashed line), the estimates of $\pe$ are upper bounds on $\pe$. The other lines in the $\pe$ plots are neither upper bound nor lower bound. The black lines in the plots, which is the estimate of $\pe$ calculated from $\smi_{MC}$, is also not the true $\pe$ but can serve as a benchmark for an upper bound on $\pe$. 
  
\begin{figure*} [htbp!]
\begin{subfigure}{0.33\textwidth}
\centering
\includegraphics[width=1\linewidth]{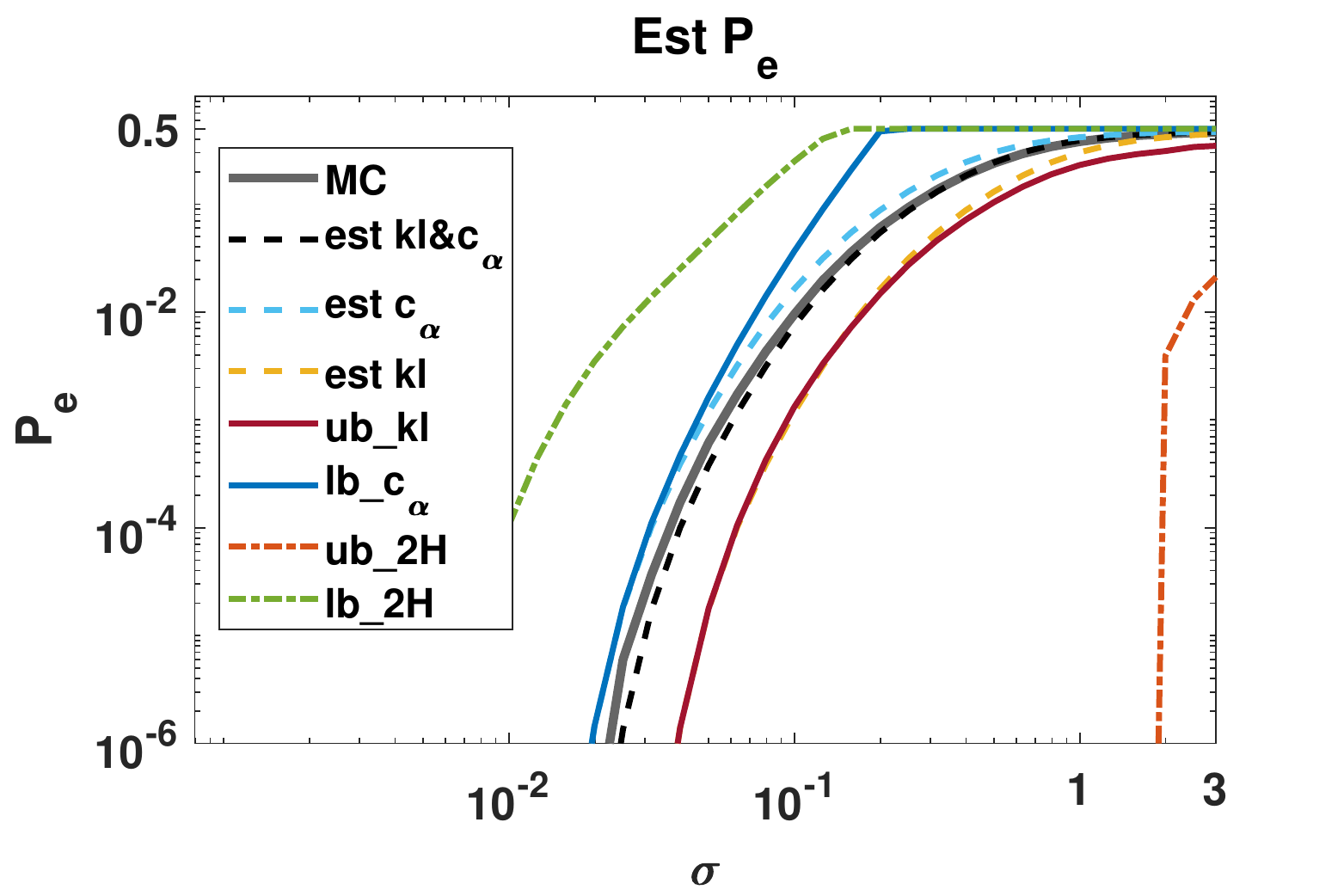}
\caption{}
\label{fig:scene1_pe}
\end{subfigure}
\begin{subfigure}{0.33\textwidth}
\centering
\includegraphics[width=1\linewidth]{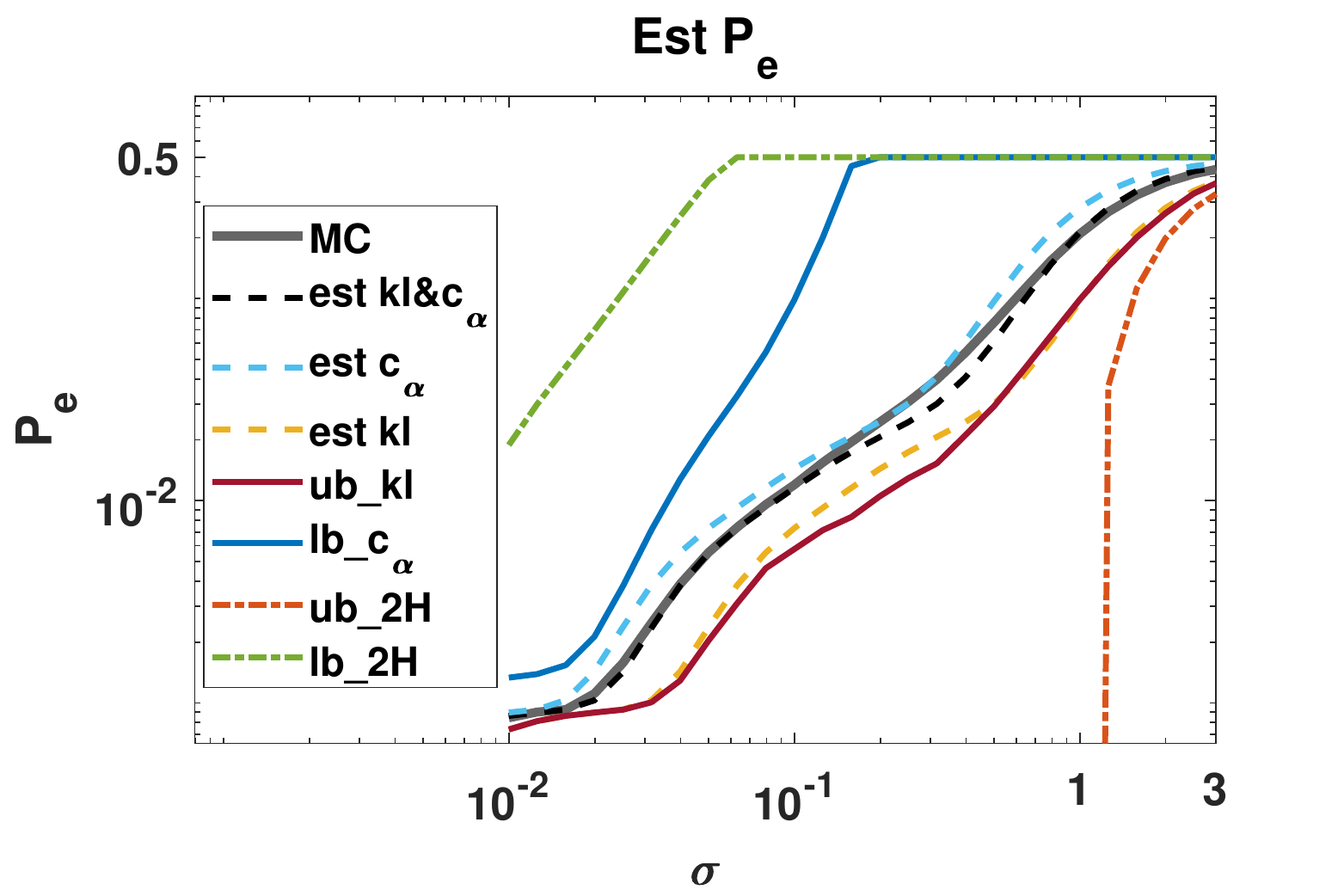}
\caption{}
\label{fig:scene2_pe}
\end{subfigure}
\begin{subfigure}{0.33\textwidth}
\centering
\includegraphics[width=1\linewidth]{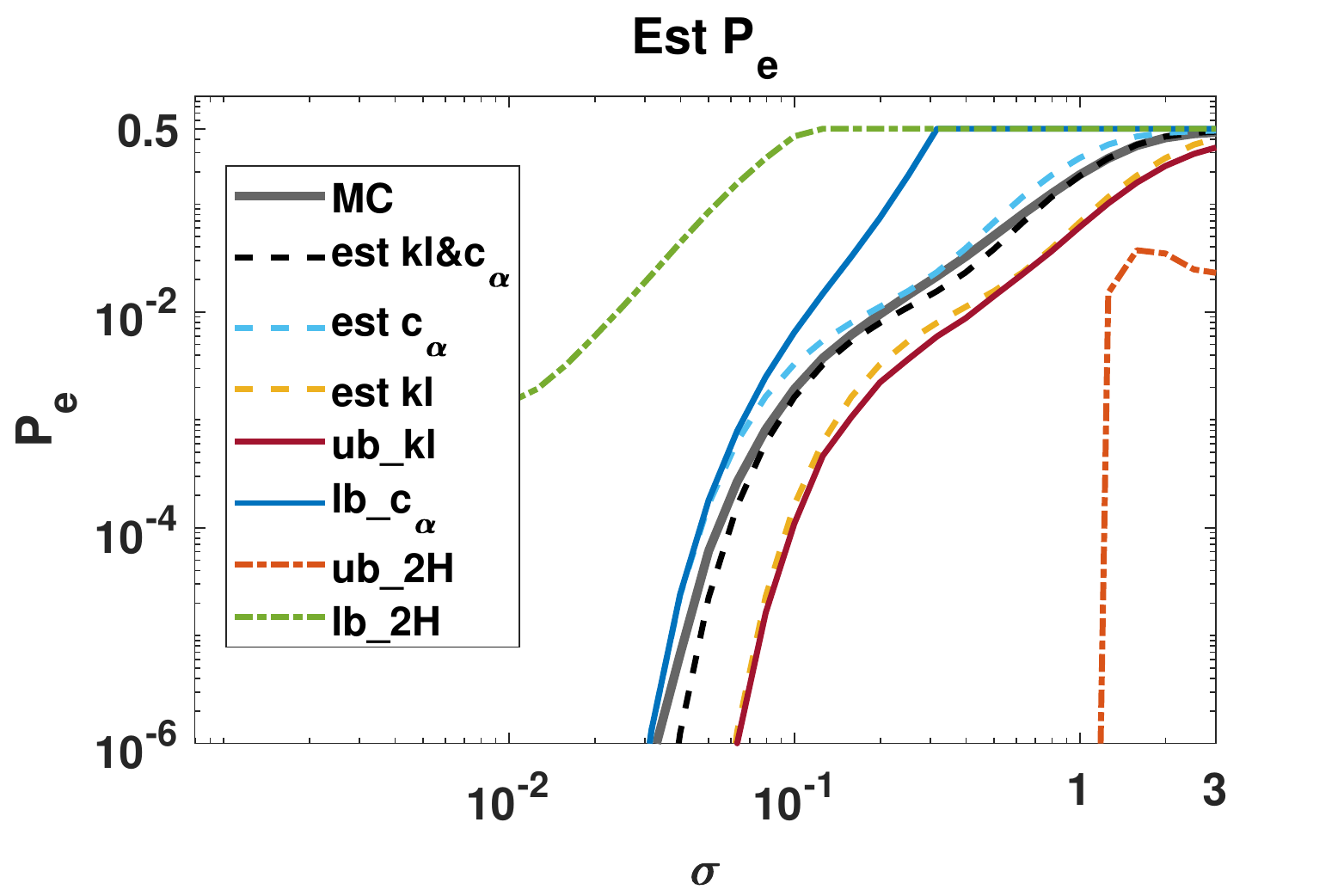}
\caption{}
\label{fig:scene3_pe}
\end{subfigure}
\caption{Estimates of $\pe$ calculated from a number of estimators of $\mathrm{I(\bm x; C)}$. }
\label{fig:pe_plots}
\end{figure*}

\indent In Figure~\ref{fig:pe_plots}, the blue and dark red solid lines are significantly closer to the grey line than the green and orange dashed lines. The results demonstrated that our new bounds is tighter than the bounds derived from entropy bounds. Furthermore, the three estimators (blue, black and yellow dashed lines) are good surrogates for the true mutual information.

\bibliographystyle{IEEEtran}
\bibliography{math}

\end{document}